\documentclass[letter]{AA}

\usepackage[colorlinks, citecolor=blue]{hyperref}
\usepackage{pdflscape}
\usepackage{xcolor}
\usepackage{lscape}

\usepackage{graphicx}
\usepackage[normalem]{ulem}
\newcommand{\kms}{km\,${\rm s}^{-1}$}
\newcommand{\Msun}{M$_{\odot}$}

\usepackage{txfonts}

\begin{document}

   \title{
   HR\,6819 is a binary system with no black hole\thanks{Based on Director Discretionary Time (DDT) observations made with ESO Telescopes at the Paranal Observatory under programme IDs 2107.D-5026, 63.H-0080 and 073.D-0274.}
   }
   \titlerunning{HR\,6819 is a binary system with no black hole}
   \subtitle{Revisiting the source with infrared interferometry and optical integral field spectroscopy}
  

   \author{A. J. Frost\inst{1}, 
   J.\,Bodensteiner\inst{2}, 
   Th.\,Rivinius\inst{3}, 
   D.\,Baade\inst{2},
   A.\,Merand\inst{2},
   F.\,Selman\inst{3},
   M.\,Abdul-Masih\inst{3},
   G.\,Banyard\inst{1},
   E.\,Bordier\inst{1,3},
   K.\,Dsilva\inst{1},
   C.\,Hawcroft\inst{1},
   L.\,Mahy\inst{4},
   M.\,Reggiani\inst{1},  
   T.\,Shenar\inst{5}, 
   M.\,Cabezas\inst{6}, 
   P.\,Hadrava\inst{6}, 
   M.\,Heida\inst{2}, 
   R.\,Klement \inst{7}
   \and
   H.\,Sana\inst{1}
   }
   
\authorrunning{A. J. Frost et al.}
 \institute{Institute of Astronomy, KU Leuven, Celestijnlaan 200D, 3001 Leuven, Belgium\\
              \email{afrost274@gmail.com}
            \and
            European Organisation for Astronomical Research in the Southern Hemisphere (ESO), Karl-Schwarzschild-Str. 2, 85748 Garching b. München, Germany
            \and
            European Organisation for Astronomical Research in the Southern Hemisphere (ESO), Casilla 19001, Santiago 19, Chile
            \and
            Royal Observatory of Belgium, Avenue Circulaire 3, B-1180 Brussel, Belgium
            \and
             Anton Pannekoek Institute for Astronomy, University of Amsterdam, Postbus 94249, 1090 GE Amsterdam, The Netherlands 
            \and
            Astronomical Institute, Academy of Sciences of the Czech Republic, Bo\v{c}n\'{\i} II 1401, 141 31 Praha 4, Czech Republic
            \and
            The CHARA Array of Georgia State University, Mount Wilson Observatory, Mount Wilson, CA 91023, USA
             }

   \date{Received December 24, 2021; Accepted February 1, 2022.}

 
  \abstract
   {Two scenarios have been proposed to match the existing observational constraints of the object HR\,6819. The system could consist of a close inner B-type giant plus a black hole (BH) binary with an additional Be companion in a wide orbit. Alternatively, it could be a binary composed of a stripped B star and a Be star in a close orbit. Either scenario makes HR\,6819 a cornerstone object as the stellar BH closest to Earth, or as an example of an important transitional, non-equilibrium phase for Be stars with solid evidence for its nature.}
   {We aim to distinguish between the two scenarios for HR\,6819. Both models predict two luminous stars but with very different angular separations and orbital motions. Therefore, the presence of bright sources in the 1-100~milliarcsec (mas) regime is a key diagnostic for determining the nature of the HR\,6819 system.}
   {We obtained new high-angular resolution data with VLT/MUSE and VLTI/GRAVITY of HR 6819. The MUSE data are sensitive to bright companions at large scales, whilst the interferometric GRAVITY data are sensitive down to separations on mas scales and large magnitude differences.}
   {The MUSE observations reveal no bright companion at large separations and the GRAVITY observations indicate the presence of a stellar companion at an angular separation of $\sim1.2$ mas that moves on the plane of the sky over a timescale compatible with the known spectroscopic 40-day period. 
   }
   {
   We conclude that HR\,6819 is a binary system and that no BH is present in the system. The unique nature of HR\,6819, and its proximity to Earth make it an ideal system for quantitatively characterising the immediate outcome of binary interaction and probing how Be stars form.}


   \maketitle
%

\section{Introduction} \label{s:INTRO}


The majority of main-sequence (MS) massive OB-type stars belong to binary or higher-order multiple systems, among which  
close binaries ($P$ < 10\,yr) are common \citep{mason98, sana12, sota14, moe17, Almeida2017, Banyard2021, Villasenor2021}. Close companions will most likely interact and exchange mass and angular momentum, significantly impacting their subsequent evolutionary path and final fates. Binary interactions are likely to leave the involved stars chemically and physically altered \citep{pac71, pols91, demink13}, although high-quality observations, which theoretical predictions might be compared to, are lacking \citep{laurent20b,laurent20a}. If the central core is massive enough and the system survives both the interaction and a potential supernova explosion, it may result in a MS star and compact object binary \citep{langer2020}. Those are prime candidate progenitors of high- and low-mass X-ray binaries (e.g. \citealt{liu06}), and for a small fraction of them, of double compact binaries and gravitational wave (GW) mergers (e.g. \citealt{Abbott2019}). 

HR\,6819 (also known as HD~167\,128, ALS~15\,056, and QV~Tel) is a highly intriguing object that was recently proposed as a candidate multiple system containing a stellar-mass black hole (BH). Based on optical spectroscopic monitoring over $\sim$5yr, \citet{rivi20} determined HR\,6819 to be a narrow, singled-lined spectroscopic binary with a 40 d orbit and a negligible eccentricity. The authors proposed the HR\,6819 system to be a hierarchical triple, with a distant classical Be star orbiting a short-period inner binary consisting of a B3~III star and a stellar, quiescent (non-accreting) BH in a close circular orbit. Classical Be stars are rapidly rotating B stars with decretion disks that produce strong emission lines (e.g. \citealt{rivi2013}), with studies suggesting that they may form as a result of mass and angular momentum transfer in binary systems (e.g. \citealt{wang21,julia2020b,klem19}). This would make HR\,6819 a lower mass counterpart of LB-1, another intriguing spectroscopic binary system \citep{irr20}. 
While early works implied that the detected RV signal was spurious and a more conservative mass BH could reproduce the observations of LB-1 \citep{lb1, elbad1, simondiaz20}, \citet{tomer} showed that the spectral variations could instead be due to a rare Be binary system consisting of a stripped star and a Be star rotating near its critical velocity. Both models were additionally tested by \citet{lennon21}.

For HR\,6819, \citet{juliahr6819} and \citet{elbad2} independently proposed that the spectral observations could similarly be caused by a binary system consisting of a stripped B-type primary and a rapidly-rotating Be star that formed from a previous mass-transfer event. 
Additionally, \citet{Gies2020} reported a small reflex motion detected in the H$\alpha$ line arising in the Be star disk, invoked by a companion in a close orbit, therefore yielding the same conclusion. In the meantime, \citet{Safarzadeh2020} proposed based on stability arguments that the triple configuration is highly unlikely. On the other hand, \citet{Mazeh2020} argued that if HR\,6819 is indeed a triple system, the putative BH could itself be an undetected binary system of two A0 stars, making HR\,6819 a quadruple system. \citet{Klement2021} later reported on speckle observations (obtained with the Zorro imager on the Gemini South telescope in Chile) that indicated a possible optical source at 120\,mas from the central source. Given that the brightness of the source could not be fully constrained (it can be up to five magnitudes fainter than the central object), it could either be the Be star and thus an indication for the triple scenario, or an unrelated background or foreground source.

The main difference between the two main proposed scenarios is the separation of the two luminous sources. The triple scenario relies on a wide orbit for the two bright stars in the system with an angular separation of $\sim$100\,mas, so that significant shifts are not generated in the RV measurements over five years. On the other hand, the binary model requires a much smaller separation of $\sim$1-2\,mas.
Both models have consequential implications. If HR\,6819 were to contain a BH, the system would constitute a prime testing centre for investigating supernova kicks and GW-progenitors. If HR\,6819 were to contain a stripped star, it would provide a direct observation of a binary system briefly after mass transfer as well as evidence that the binary channel is a way in which classical Be stars form.

In this letter we present new observations of HR\,6819 designed to distinguish between these two scenarios. Specifically, observations from the GRAVITY and Multi Unit Spectroscopic Explorer (MUSE) instruments at the Very Large Telescope/Interferometer (VLT/I) were selected. The MUSE data are sensitive to a wide companion, with separations in the range of $\sim$100\,mas to 7.5", whilst the GRAVITY data cover the range from a fraction of a mas to $\sim$100\,mas. In Sections \ref{s:MUSE} and \ref{s:GRAVITY} we present our MUSE and GRAVITY observations respectively. Our results are discussed and summarised in Section~\ref{s:ccl}, where we offer our conclusion that HR\,6819 is a close binary consisting of a stripped B star and a rapidly rotating Be star.


\section{MUSE: Search for a wide companion} \label{s:MUSE}
\subsection{Observational setup and data reduction}
MUSE is an integral field spectrograph operating over visible wavelengths (480-930nm) \citep{bacon2010}. The MUSE observations were designed to detect a possible wide companion. The spectroscopic observations implied that the B star and Be star are of similar brightness \citep{rivi20, juliahr6819}, and the speckle interferometry suggested a source at most five magnitudes fainter than the central source at a separation of 120\,mas \citep{Klement2021}. Therefore, the observations were set up to yield a signal-to-noise (S/N) of 25 in H$\alpha$ for a putative companion at 120\,mas that is five magnitudes fainter than the central source.

One epoch of VLT/MUSE observations of HR\,6819 was obtained on July 22, 2021 (2107.D-5026, PI: Rivinius) in the narrow-field mode (NFM), supported by adaptive optics (AO), and in nominal wavelength mode (N). The NFM covers a 7.5" $\times$ 7.5" field of view (FoV) with a spatial sampling of 0.025" $\times$ 0.025". At a distance of 340~pc, the FoV corresponds to $\sim$2550\,AU (or $\sim$0.12\,pc). Given the brightness of the target (V\,=\,5.4), the observations were not carried out in the normal NFM-AO mode but using a procedure that widens the laser configuration and thus allows the observations of brighter stars. The observation was split into several subsets with a small ($\sim$1") offset and position angle 0 or 90$^{\circ}$. Each set was then made up by 6 or 12 short (3-second) exposures, leading to a total of 36 exposures with an overall exposure time of 108 seconds.  

The individual MUSE NFM AO N observations were reduced using the standard MUSE pipeline \citep{musepipe}, which includes bias subtraction, flat fielding, wavelength calibration and illumination correction for each individual integral field unit (IFU). After recombining the data subsets from all the IFUs, the data were flux-calibrated using a standard star and a sky subtraction was performed. Given the extremely good weather conditions during the observations (at the start of the first exposure, the seeing was 0.58", which varied between the observations and increased to around 0.86"), most of the exposures are saturated in the central source. This does not affect the purpose of the MUSE observations, however, which is to probe whether there is a wide companion at $\sim$120\,mas. 

      \begin{figure*}
   \centering
   \includegraphics[width=185mm]{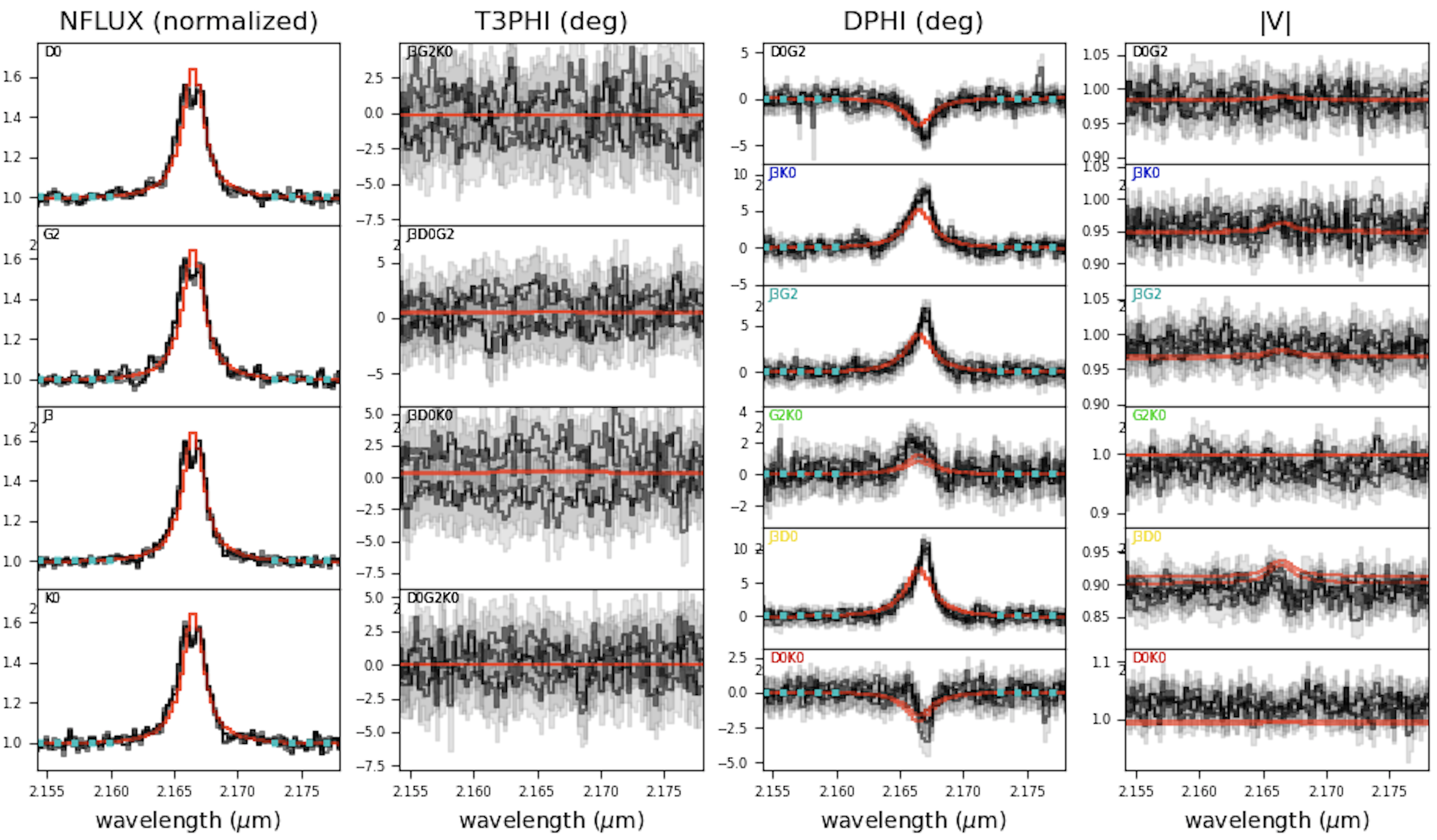}
      \caption{Best fits to the GRAVITY dataset taken on September 19 2021 formed from a model composed of two point sources. The data are shown in black and the model is shown in red over a wavelength range highlighting the Br$\gamma$ region as opposed to the whole GRAVITY wavelength range. In the appendices, we explain in detail how we determined this as our final fit to the data. While the different rows show different VLTI baselines (see inset labels), the different panels show the normalised flux, the closure phases, the differential phases and visibilities, from left to right. The blue dots represent the continuum.
              }
         \label{endbinfit}
   \end{figure*}

\subsection{Results}
The individual (not saturated) MUSE exposures show one central source. Collapsing the wavelengths to a white-light image (see Figure \ref{muse}) we find no indication for a second bright source at 120\,mas. Using the spectroscopic capabilities of MUSE, we find no indication for a second star farther out (which could for example be indicated by spatially localised absorption lines in the Ca triplet at around 8500\,\AA). Additionally, there seems to be no significant spatial change in the spectrum of the central source, so we extracted the spectrum by summing over the 3$\times$3 most central pixels with QFitsView\footnote{\url{https://www.mpe.mpg.de/~ott/QFitsView/}}. 
We compared this extracted spectrum to an observed FEROS spectrum
\citep[described in][]{rivi20} at a similar orbital phase, 
after resampling the spectra to the MUSE resolution and binning. The comparison (see Fig.\,\ref{spec_comparison}) shows that there is no significant difference between the spectra taken by the two instruments, implying that the two sources that contribute to the composite FEROS spectrum (the B and the Be star, which are common to both models) are both located within in the central source detected in the MUSE data. These two findings combined (i.e. the lack of a second bright source at 120\,mas from the central source, and the fact that both the B and the Be star both contribute to the spectrum of the central source) unambiguously show that the Be star in HR\,6819 is not a wide companion located at 120\,mas as suggested by the speckle interferometry.


   \begin{figure*}
   \centering
   \includegraphics[width=160mm]{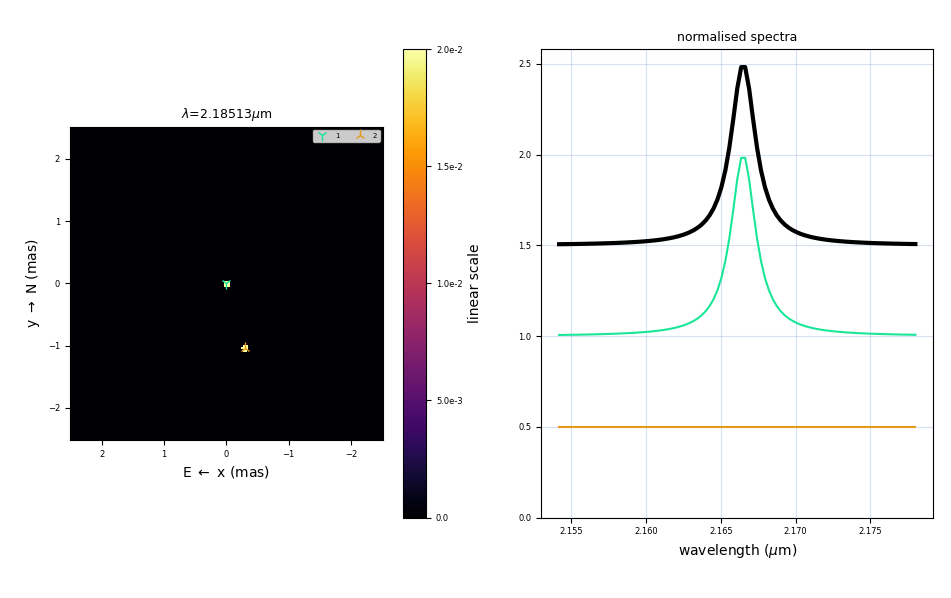}
    \caption{Model image (left) and synthetic spectra (right) corresponding to the best-fit model shown in Fig.\,\ref{endbinfit}. The green line corresponds to the primary star that is fixed at (0,0) and whose continuum flux is normalised to 1. The orange line corresponds to the secondary star. The black line is the total spectrum. The spectral line included in the model is at the Br$\gamma$ wavelength.
              }
         \label{endbinmod}
   \end{figure*}

\section{GRAVITY: Probing the inner regions} \label{s:GRAVITY}


\subsection{Observations and fitting process}
GRAVITY is the $K$-band four-telescope beam combiner at the Very Large Telescope Interferometer \citep{gravity}. Three GRAVITY observations of HR\,6819 were obtained in August and September 2021 using the high spectral resolving power setting of GRAVITY ($R=\lambda / \Delta \lambda =4000$). Observations were taken with the large configuration (A0-G1-J2-J3) of the 1.8~m Auxiliary Telescopes (ATs). The first observation (August 2021) was unsuccessful, as severe weather and technical issues only allowed the science source to be observed with no calibrator. 
The following two attempts, on September 6 and 19 2021, were successful, however, with clear sky conditions, precipitable water vapour below 30\,mm, and seeing of less than 1" for both the science source and the calibrator star (HD~161420). The data were reduced and calibrated using the standard GRAVITY pipeline \citep{gravpipe}.

Geometric fitting was applied to the interferometric observables (i.e., visibilities, closure phases, differential phases and normalised fluxes) 
using the Python3 module PMOIRED\footnote{\url{https://github.com/amerand/PMOIRED}}, which allows the display and modelling of interferometric data stored in the OIFITS format. 
Notably, a strong Br$\gamma$ line is visible across the observables. When fitting models with PMOIRED, the primary is described as the central source, fixed at position (0,0) in the FoV. The flux of the primary star is normalised to 1 so it can be used as a reference point to determine the relative fluxes of any companions. Additionally, any present emission lines in the fluxes and visible in the differential phases can be modelled, by attributing Lorentzian or Gaussian line profiles to the model components. The fitting process was judged using the reduced chi-square statistic.  

\subsection{Results}

A variety of models were run, including single star models, single disk models, binary disk models and triple systems. We refer to Appendix \ref{appendix} for an in-depth description of the fitting process. The best fit to the GRAVITY data, with an example shown in Fig.\,\ref{endbinfit}, comes from a model composed of two sources, implying that a binary system with two optically bright companions is present at GRAVITY scales at a separation of $\sim$1\,mas across both epochs. 
Table \ref{params} displays the specific parameters resulting in the best-fitting model. We determined the errors on our derived measurements of the sources using bootstrapping. 
In the bootstrapping procedure, data are drawn randomly to create new datasets 200 times and the final parameters and uncertainties are estimated as the average and standard deviation of all the fits that were performed. 
Figure \ref{endbinmod} displays a model image and the synthetic spectra of the model components. The dimmer star is on average 56\% the flux of the brighter star on average and in this best-fitting model the Br$\gamma$ emission comes exclusively from the brightest star in the model.



   \begin{figure}
   \centering
   \includegraphics[width=85mm]{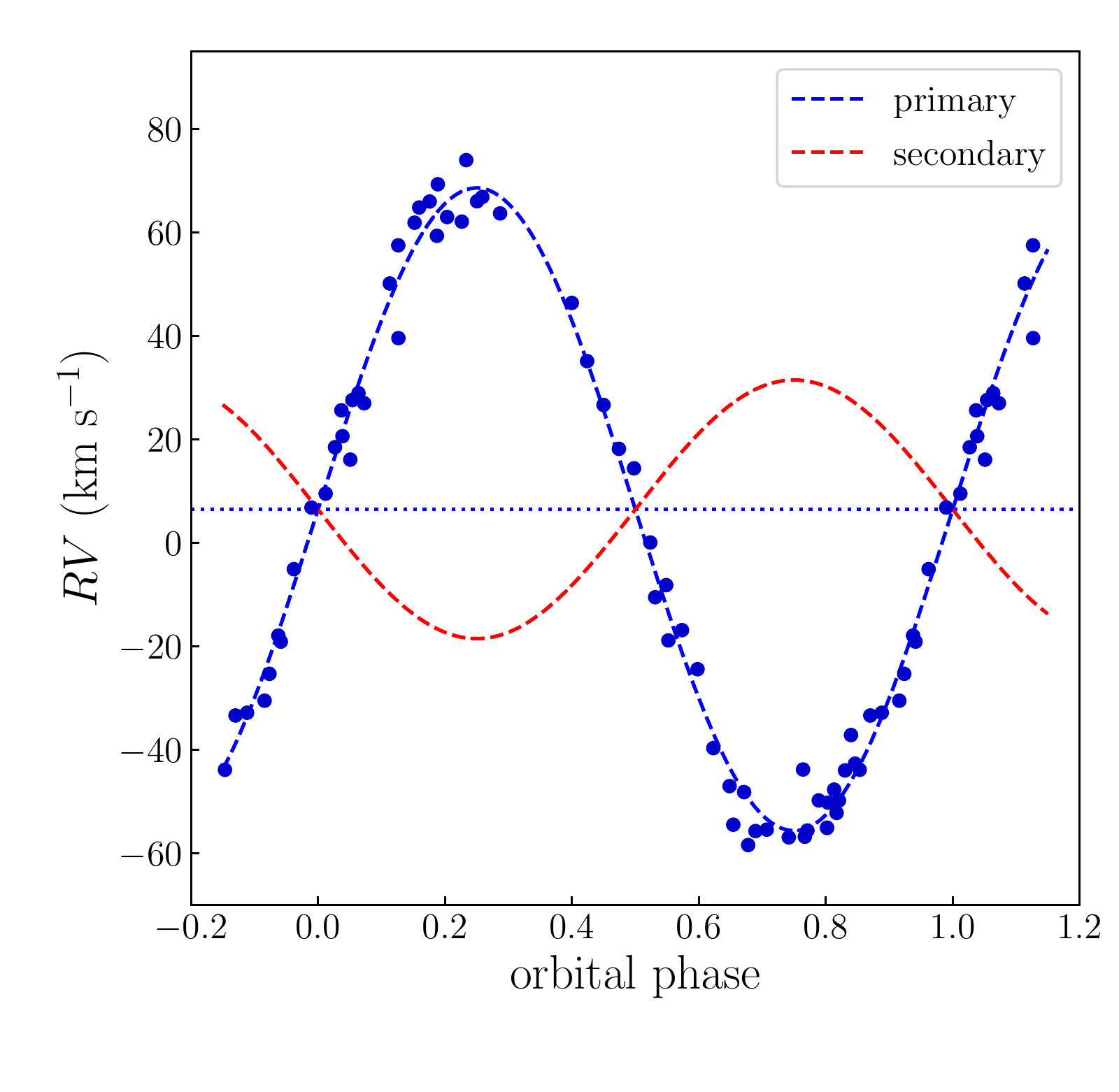}
   \includegraphics[width=85mm]{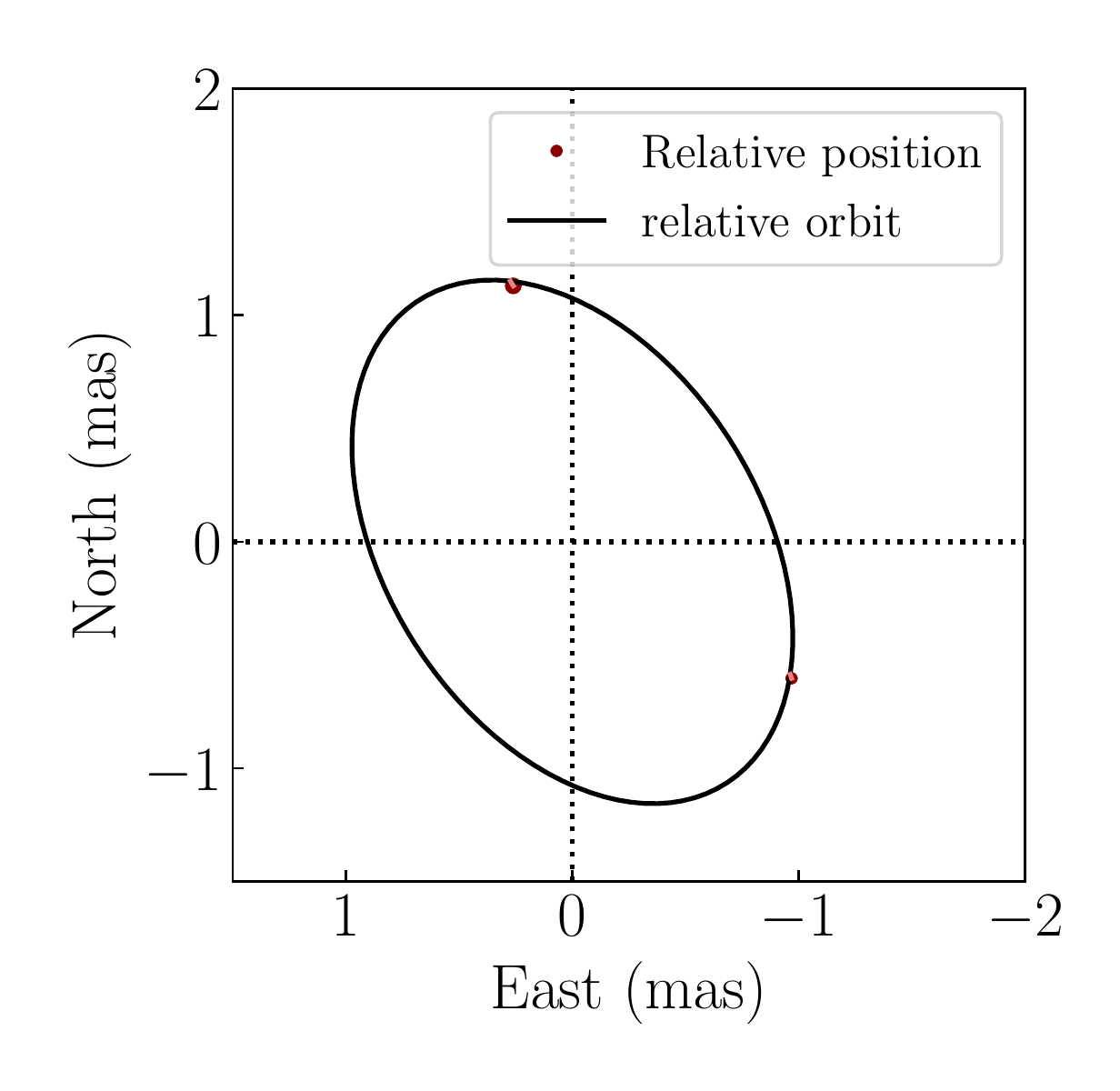}
      \caption{Simultaneous fit to the FEROS RVs of the stripped B star (upper panel) and the GRAVITY relative astrometry (lower panel) obtained using an adopted distance of 340~pc (Table~\ref{orbitparams}). The corresponding MCMC grid is displayed in Fig.~\ref{mcmc}. The dashed blue and red lines in the top panel show the best-fit RV of the stripped star and the predicted secondary RV-curve for the adopted solution respectively (see other solutions discussed in the text). 
      The horizontal dotted line shows the systemic velocity of the system. In the bottom panel, we followed the definition of \citet{juliahr6819} for the identification of the primary and secondary components in the system and placed the suspected stripped B star at coordinate (0,0).  } 
         \label{orbit}
   \end{figure}

\subsection{Astrometric orbit}

The interferometric data provide us with astrometric information about the luminous stars in the system. We fed this astrometric data into the SPectroscopic and INterferometric Orbital Solution finder (\textsc{spinOS}, \citealt{spinos}) to compare the movement of the system as constrained by interferometry to the orbit found by \citet{juliahr6819}. 
Given the limited astrometric data, we opted to to fix the distance. \citet{rivi20} suggested a distance of $310\pm60$~pc while Gaia yielded values of 342$_{-21}^{+25}$\,pc  and 364$_{-16}^{+17}$\,pc in DR2 and eDR3, respectively \citep{dr2,dr3}. It is unclear, however, how the presence of two optical components in HR~6819 impacts the Gaia distance estimate. Astrometric excess noise can be caused by the high target brightness as well as by binarity and thus the purely statistical errors given by the parallax error parameter of Gaia can be incomplete. For source 6649357561810851328 (HR 6819) the astrometric excess noise is marked as 0.857 mas for eDR3 at a significance level of 1180 and 0.731 mas at a significance level of 247 for DR2 and thus these two values overlap within their error ranges. We adopted a distance value of 340~pc, which sits comfortably within the estimated range of distances and is consistent with the distance value adopted in \citet{juliahr6819}.

Furthermore, we used the presence of the strong Br$\gamma$ emission to identify the relative locations of the Be and B stars in the GRAVITY data, as the best interferometric fit determines that the emission determines from the brighter star only. We also used the orbital parameters ($P, i, T_0$) of \citet{juliahr6819} as initial guesses as well as the measured RVs of the narrow-line star to perform a first optimisation of the combined astrometric and RV data  through the Levenberg-Marquardt optimiser in \textsc{spinOS}. The orbit was assumed to be circular. The obtained parameters then served as input for a full Markov Chain Monte Carlo (MCMC) optimisation where the period, inclination, time of perihelion passage, ascending node, semi-amplitude and total mass ($P, i, T_0, \Omega, K_1, \gamma_1$, and $M_\mathrm{tot}$, respectively) were varied simultaneously. The resulting best-fit astrometric orbit and RV curve are shown in Fig.~\ref{orbit}.  

The new astrometric data are simultaneously fit together with the RV measurements of the narrow-line star with a fixed distance of 340~pc. We recovered the orbital parameters of \citet{juliahr6819}, except for a higher inclination of $49\pm2$\degr\, (compared to their estimated value of $\sim$32\degr) and a small adjustment of the ephemeris within the uncertainty of \citet{juliahr6819} (Table~\ref{orbitparams}). Our combined solution yields a total mass for the system of 6.5$\pm$0.3\,M$_{\odot}$, with the errors estimated from the MCMC (Fig.~\ref{mcmc}). This derived mass is in excellent agreement with the masses estimated from the atmospheric analysis of the Be and stripped B star scenario of \citet{juliahr6819}. However, the estimated mass ratio would yield $K_2\approx 25$~\kms, in stark contrast with the value of $4\pm1$~\kms estimated from spectral disentangling. An inclination of 35\degr\ would be needed to reconcile the two $K_2$ values. A solution with an inclination fixed at $i=37$\degr is possible, but the residuals of the astrometric solution increase from 11$\mu$as to 67$\mu$as. In this case, the derived total mass is 5.8$\pm 0.2$M$_{\odot}$. Possible further avenues to reconcile the two datasets include a much closer distance ($\sim 260$~pc, Table~\ref{orbitparams}), larger statistical errors on the GRAVITY measurements, small systematic errors ($\sim$70\,$\mu$as) in the relative positions listed in Table~\ref{params}, or a combination of these.

The assumed distance indeed has a strong impact on the estimated total mass (Table~\ref{orbitparams}). Varying the distance by $+25$~pc (or $-30$~pc)  translates into a $\pm 1.5$~\Msun\ shift of the total mass. In addition,  astrometric residuals of the fit of around 10\,$\mu$as clearly indicate that we are over-fitting the data, that is we are lacking a sufficient number of observational constraints with respect to the number of model parameters. Uncertainties, in this situation, are typically underestimated and the goodness-of-the-fit cannot be used as an estimate of the quality of the model. It is therefore clear that more GRAVITY data are required if reliable estimates of the orbit orientation and the individual component masses of HR~6819 are to be obtained. More interferometric data would also help to lift the uncertainties surrounding the Gaia estimate. Despite these limitations, the fact that the MUSE data described in the previous section detect no wide, bright companion, and the fact that GRAVITY shows two bright objects in a 40-day orbit lead us to conclude that the (B+BH)+Be triple system scenario should be excluded at a high confidence level.

\section{Conclusions} \label{s:ccl}

We have presented new MUSE and GRAVITY data for the exotic source HR\,6819 in order to distinguish between the two proposed hypotheses for the nature of the system. The first scenario suggests that HR\,6819 is a triple system with an inner B star plus BH binary and an outer Be star tertiary companion on a wide orbit \citep{rivi20} and the second scenario suggests that the system is a binary system consisting of a Be star and a B star that have previously interacted \citep{juliahr6819, elbad2}. If the first scenario were correct, a bright quasi-stationary companion should be present at $\sim$100\,mas. If the second scenario were correct, no bright companion should be detected at this separation and two bright stars should be detected at small separation.

Our MUSE data show no bright companion at $\sim$100\,mas separation. Our GRAVITY data resolve a close binary with $\sim$1\,mas separation composed of two bright stars. The GRAVITY spectra show the characteristic Br$\gamma$ emission associated with a Be star decretion disk \citep{rivi2013}. 
Therefore, we conclude that HR\,6819 is a binary system and reject the presence of a BH on a short-period orbit in this system. HR\,6819 therefore constitutes a perfect source for investigating the origin of Be stars and their possible formation through a binary channel. In future work, further monitoring of the system with GRAVITY will be crucial. Not only can the orbit be better constrained, but these measurements will provide distance and precise mass estimates of what is likely a newly post-interaction, bloated stripped object and its associated Be star for the first time. Together with higher-resolution spectroscopy (e.g. from UVES), abundances of both stars could be derived. With this information, HR\,6819 would constitute a corner-stone object for comparing binary evolution models.

%
\begin{acknowledgements}
    We thank  Dr. Jes\'us Ma\'iz Appell\'aniz for his helpful and thoughtful comments as referee. This research has received funding from the European Research Council (ERC) under the European Union’s Horizon 2020 research and innovation programme (grant agreement number 772225: MULTIPLES). JB and MH are supported by an ESO fellowship. TS acknowledges funding received from the European Union’s Horizon 2020 under the Marie Skłodowska-Curie grant agreement No. 101024605. We also thank the team behind QFitsView for developing their tool which allowed easy handling of our MUSE data. 
\end{acknowledgements}

\bibliographystyle{aa}

\bibliography{hr}

\appendix
\section{Investigating the MUSE data}\label{museap}

\begin{figure}
   \centering
   \includegraphics[width=0.9\hsize]{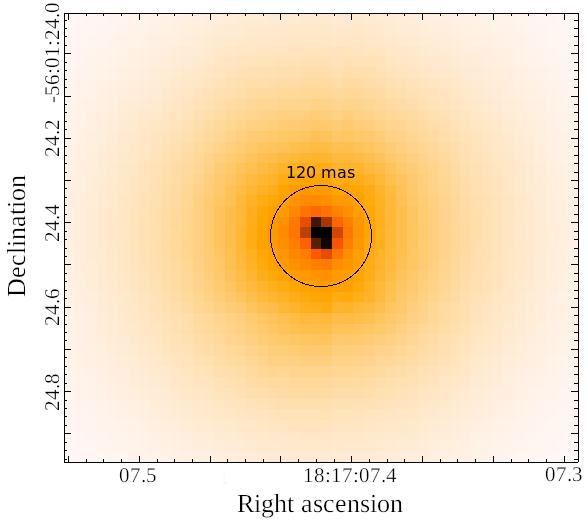}
      \caption{Cut-out around the central source in the white-light image created by collapsing one of the unsaturated MUSE exposures.}
         \label{muse}
   \end{figure}

Figure\,\ref{muse} shows a cut-out around the central source in one the MUSE exposures of HR~6819 that was created by collapsing the wavelength information in the MUSE data cube to obtain a white-light image.

\begin{figure}
   \centering
   \includegraphics[width=0.99\hsize]{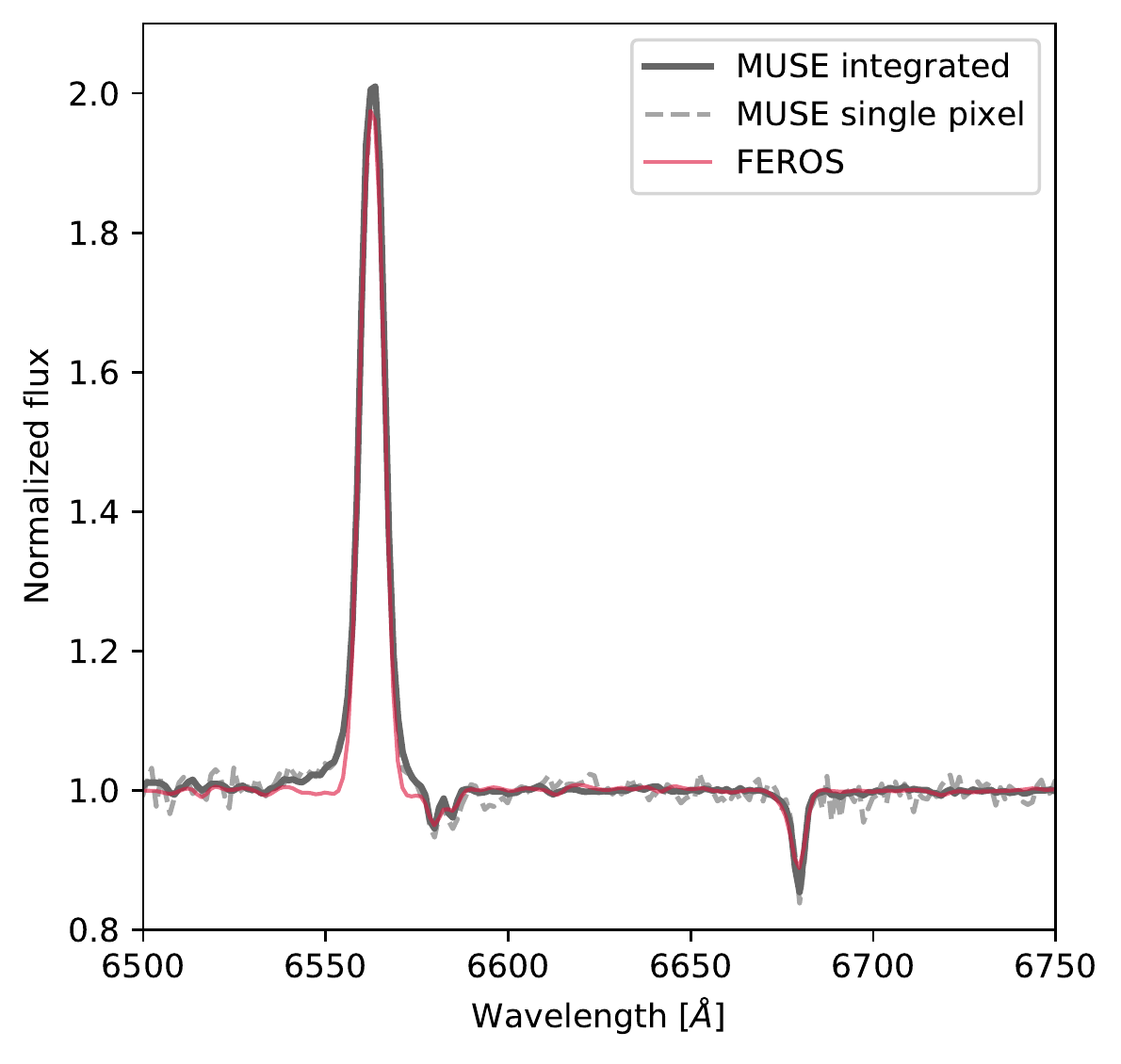}
      \caption{Comparison of two MUSE spectra of the central source, obtained by summing the 5x5 central pixels (black) or extracting only the central pixel (grey), with a downgraded FEROS spectrum (red) at a similar phase.}
         \label{spec_comparison}
   \end{figure}

Figure\,\ref{spec_comparison} compares two extracted MUSE spectra with one epoch of FEROS observations (obtained at MJD = 53248.02\,d). While one spectrum is extracted by summing the central 5x5 pixel of the source (and therefore increasing the S/N), the second spectrum is extracted just from the central pixel alone. The FEROS spectrum is downgraded to MUSE resolution and binning in wavelength. While slight differences are apparent, the overall spectra are quite similar and therefore indicate that the two stars making up the composite FEROS spectrum are within the central source that is  visible in the MUSE data.

\section{Modelling the GRAVITY data} \label{appendix}

Throughout the fitting process geometrical models were fit to the normalised K-band flux, the closure phases, the differential phases and the visibilities of the GRAVITY data using the code PMOIRED. The interferometric phases can tell us about asymmetries and the orientation of a system, whilst the visibilities describe the spatial extent of a source, with lower visibilities corresponding to a more resolved object \citep{mill15}. A perfect cosine wave pattern should be generated in the $u-v$ plane for two point sources and the deviation of the observations from this is indicative of the position angle between the sources, the frequency of the cosine is related to the separation of the sources and the contrast of the wave pattern is indicative of the flux ratio.

Moreover, GRAVITY provides spectrally resolved data, and the pattern is wavelength dependent (with a known dependence), so that in the end the available measurements very strongly over-constrain the system parameters (separation, PA and flux ratio). The GRAVITY data were taken in two polarisation planes, and the closure phase flips between these two datasets. Therefore, in order to reliably constrain the orientation of our system, the differential phase was of great importance. In our datasets a Br$\gamma$ emission line is clearly visible in the normalised flux and in the differential phase. A double peak is present in this line, particularly in the second epoch of the data (see Fig \ref{endbinfit}). This profile is often indicative of a rotating elongated structure such as a disk (e.g. \citealt{rivi2013}). Reproducing the Br$\gamma$ line proved very useful during the fitting process in order to determine the nature of HR~6819, and we describe this process in this appendix. 

A single-star model provided a poor fit to the datasets, as did a triple. Using the \texttt{detectionLimit} method of PMOIRED (the same as the approach implemented in the CANDID\footnote{\url{https://github.com/amerand/CANDID}} software \citealt{candid}) on our GRAVITY data, we find that a third companion within 100 mas of the binary must have a flux contrast of 2.0\% (2.8\%) at most in the K band compared to the primary, otherwise we would have detected it at the 3 $\sigma$ (5 $\sigma$) level.

Binary models proved much more successful. One such model that presented a good fit was a binary model in which an emission line was associated with each star. The majority of the emission came from the brightest star in the system and was emitted at the central wavelength of Br$\gamma$. The remaining contribution to the line seen in the total flux came from the dimmer star. The emission of the dimmer star in this model is slightly offset to shorter wavelengths, reproducing the double peak in the total flux. Br$\gamma$ emission coming from the two stars in a Be post-interaction binary might be plausible if some of the accreting material had failed to leave its Roche lobe and had created a small disk around the B star (as suggested in studies such as \citealt{tomer} and \citealt{levin}). However, if this were the case, the emission of the two stars would be seen to switch throughout the orbit, that is, the wavelength of the weaker emission line should change. This is not observed between the two epochs. Additionally, the superposition of two emission lines from each star could only reproduce the peak in the normalised flux, not the shape of the differential phase. Further modelling of disks around the B stars in stripped B plus Be binaries and the generation of synthetic interferometric observables could help to better quantify the effects of this potential emission.

\begin{figure}
   \centering
   \includegraphics[width=90mm]{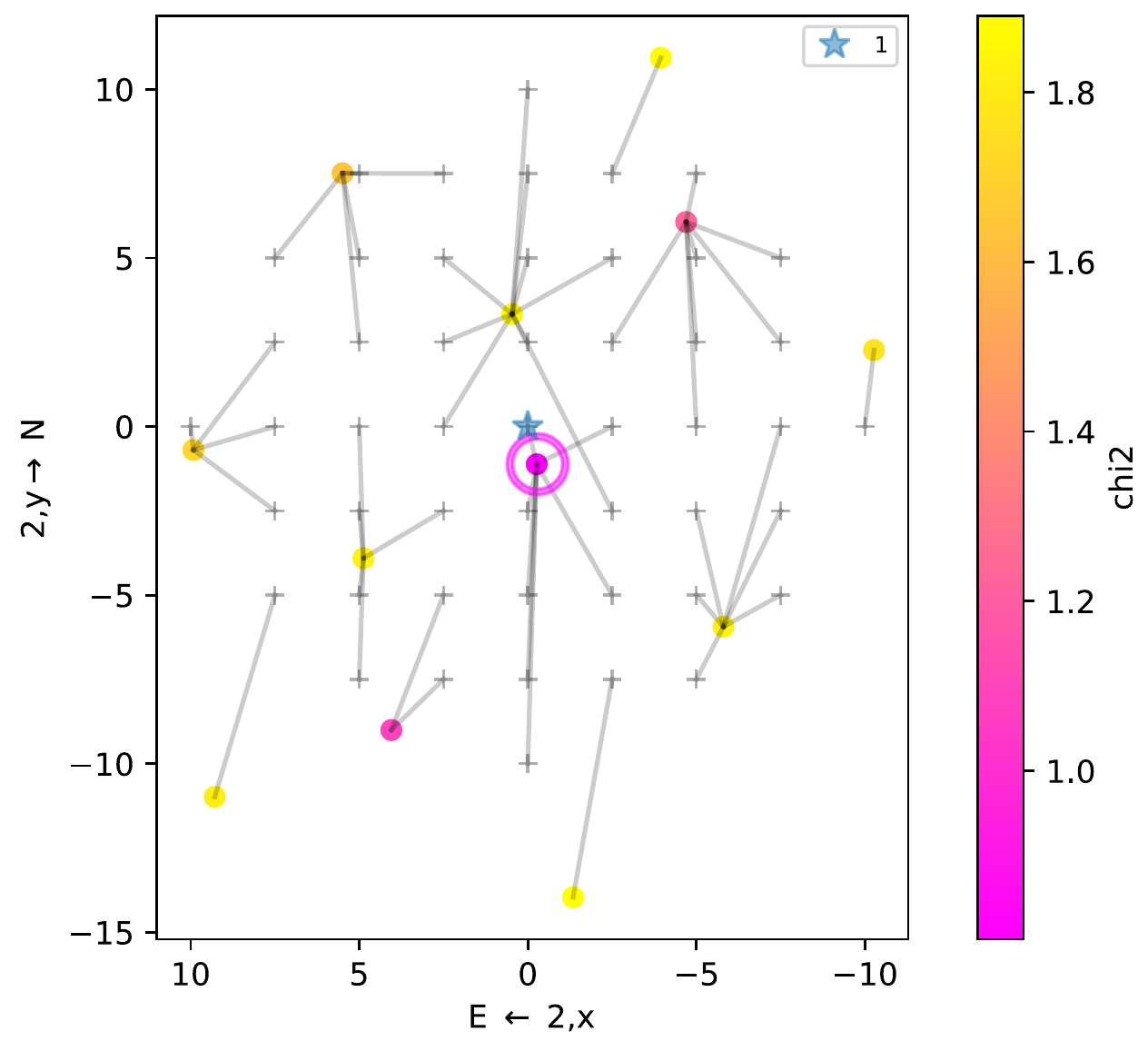}
      \caption{Grid-fit solution to the best-fitting model for a binary system with an emission line in one star. The figure shows the spatial positions that were mapped by the grid search and the associated reduced chi-square value of each of the tested positions (in colour). The central star (whose position is not fit) is shown in blue, and the axes depict the space on the sky in arcseconds. 
              }
         \label{endgrid}
   \end{figure}

Because a model with two emission lines did not produce consistent results, we ran a fit with two point sources of which only one has an emission line. This resulting fit for this model shows that the Br$\gamma$ emission comes from the brighter star in the binary and provides an excellent fit to the data as quantified by the chi-square value. PMOIRED can also run a grid search to find the best model, based on the capabilities of an associated code, CANDID \citep{candid}. Using this grid search on our datasets, the code determines that a two-point source model is the best option for the datasets, with an emission line in the brighter star. Figure \ref{endgrid} shows the result of this grid search. However, this model fails to reproduce the small double-peak feature (see Fig.\,\ref{endbinfit}). While the main peak seen in the differential phase is reproduced, the slight absorption feature is not. Despite the ambiguity, this implies that the brighter star in the system is most likely the Be star because Be star decretion disks are known to generate a surplus of Br$\gamma$ emission and infrared excess \citep{rivi2013}.

The shape of the differential phase can provide key information about a system as it is a measure of the spatial offset of the line-emitting region with respect to the continuum emission \citep{mill15, weig07}. For example, a V-shaped differential phase can imply asymmetry or simple displacement of the photocentre, whilst an S-shaped phase can imply rotation. Such S-shapes are typical of disks, and \citet{mei12} note their presence in their spectro-interferometric VLTI/AMBER survey of Be stars. In our differential phases we observe a combination of these two features. Thus, we decided to determine whether a disk model could fit the observables. Uniform disk models were first attempted, but least-square fitting was unable to converge. Be stars are expected to have Keplerian rotating disks \citep{rivi06}, therefore we also tried this, but while it was more successful the code also struggled to converge to a physical solution. This could be due to the fact that the spatial extent of the disk is too small to be resolved by GRAVITY ($<$1\,mas) whilst its emission persists in the total spectrum. 

\begin{figure*}
   \centering
   \includegraphics[width=180mm]{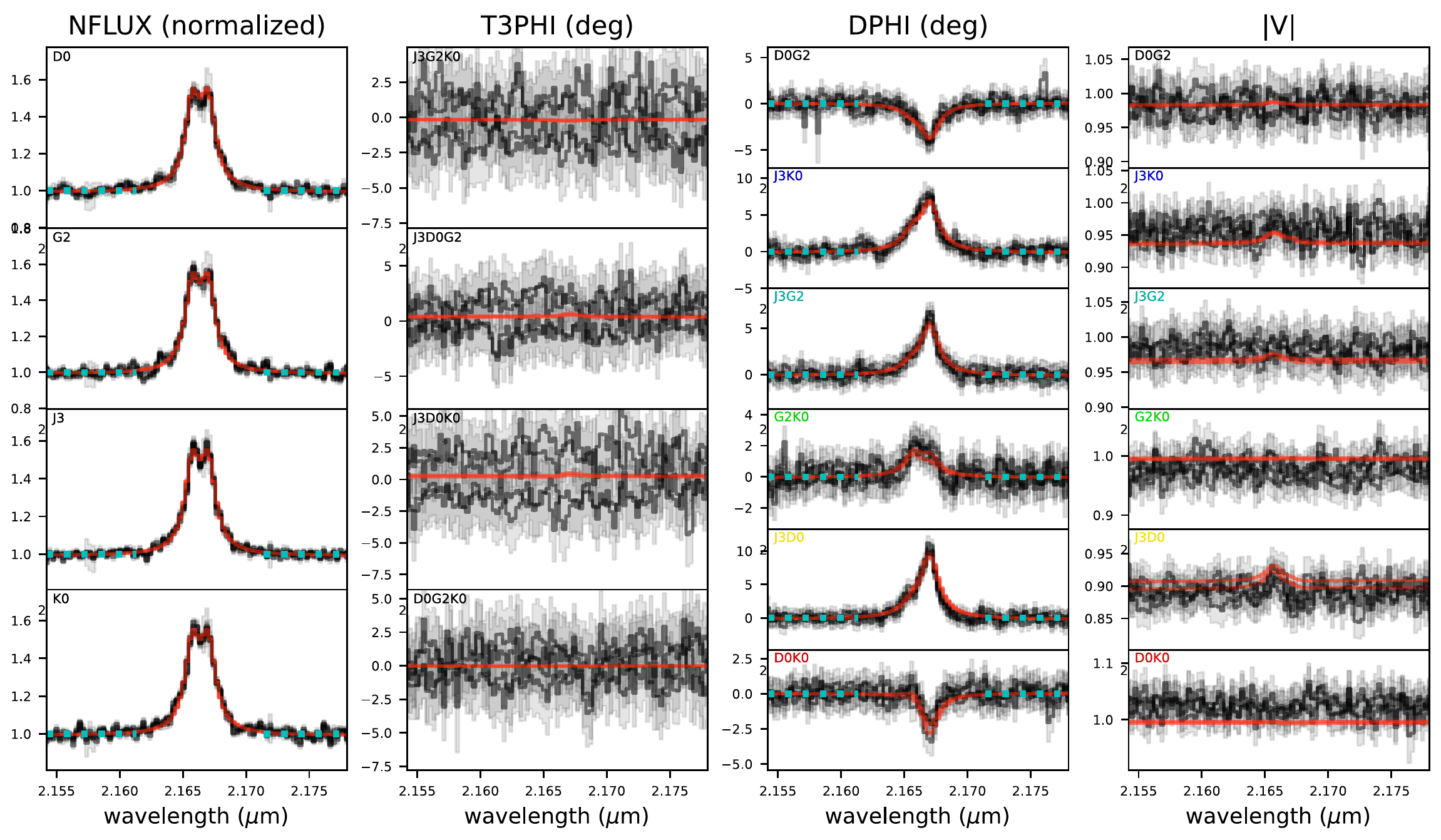}
      \caption{Fits to the second GRAVITY dataset formed by a toy model composed of two point sources and two wavelength components surrounding the primary star (meant to mimic the blue- and redshifted emission of a disk) that was run to obtain first guesses for the Keplerian fit. The inset labels are the baselines of the observations. NFLUX, T3PHI, DPHI and |V| refer to the normalised flux, closure phase, differential phase and visibilities respectively.
              }
         \label{endtoyfit}
   \end{figure*}

\begin{figure*}
   \centering
   \includegraphics[width=180mm]{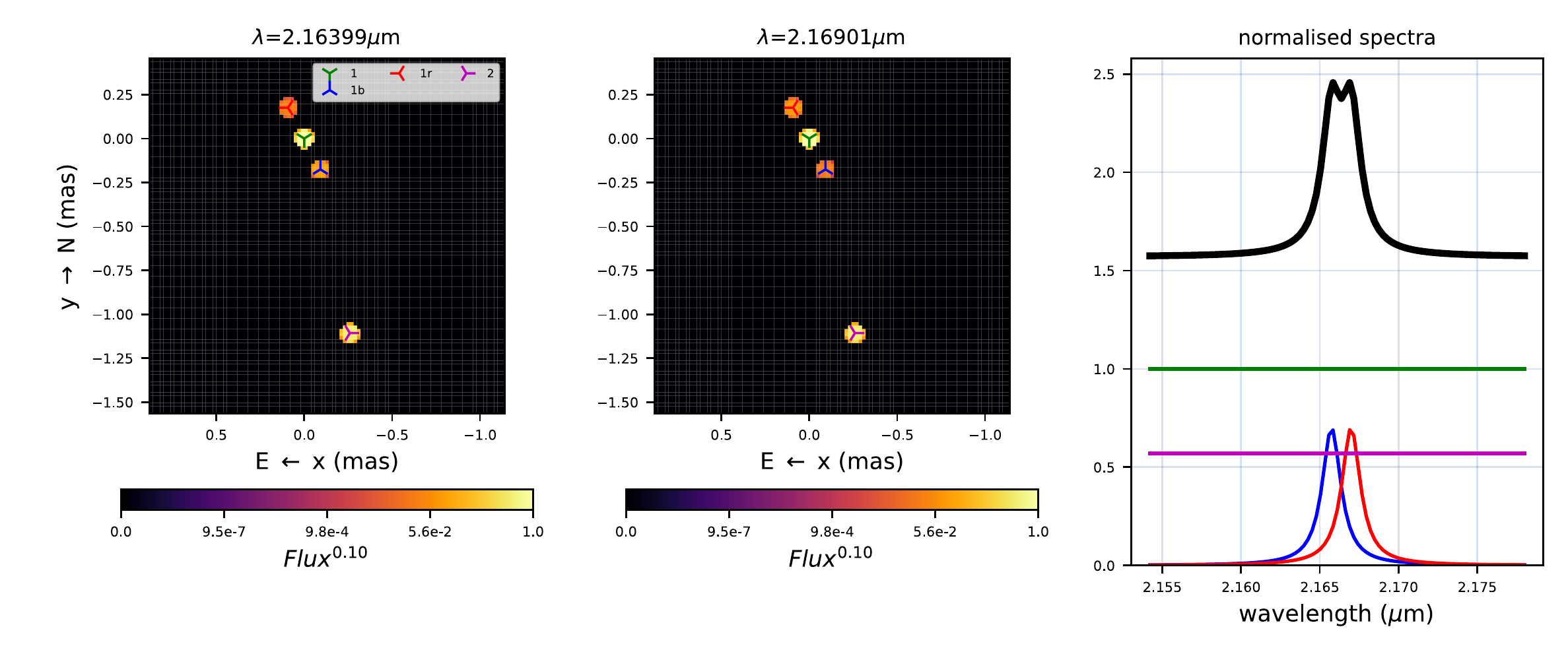}
      \caption{Model image and synthetic spectra for the toy model fit. The black line in the spectra is the total flux, the green line is the continuum flux of the brighter star, the purple line is the continuum flux of the dimmer star and the red- and blue-shifted emission is represented by the two remaining lines.
              }
         \label{endtoymod}
   \end{figure*}

A toy model was used to determine whether the blue- and redshifted emission could be used to help constrain the Keplerian fit. The model designed to represent the emission of a Be star consisted of one point source with an associated emission line at the redshifted wavelength, one point source with an emission line at the blueshifted wavelength, and one source with no emission line meant to represent the centre of the disk. The dimmer star was modelled as a point source without an emission line. Figures \ref{endtoyfit} and \ref{endtoymod} show the best-fit dataset and a visualisation of the model. The model provides a good fit to the data, particularly the second epoch, despite its nonphysical nature. The only caveat is that the fit to the visibilities of the first dataset is worse, but this is consistent because the separation of the blue and red shifted emission is not-technically resolvable with the VLTI. 

\begin{figure*}
   \centering
   \includegraphics[width=180mm]{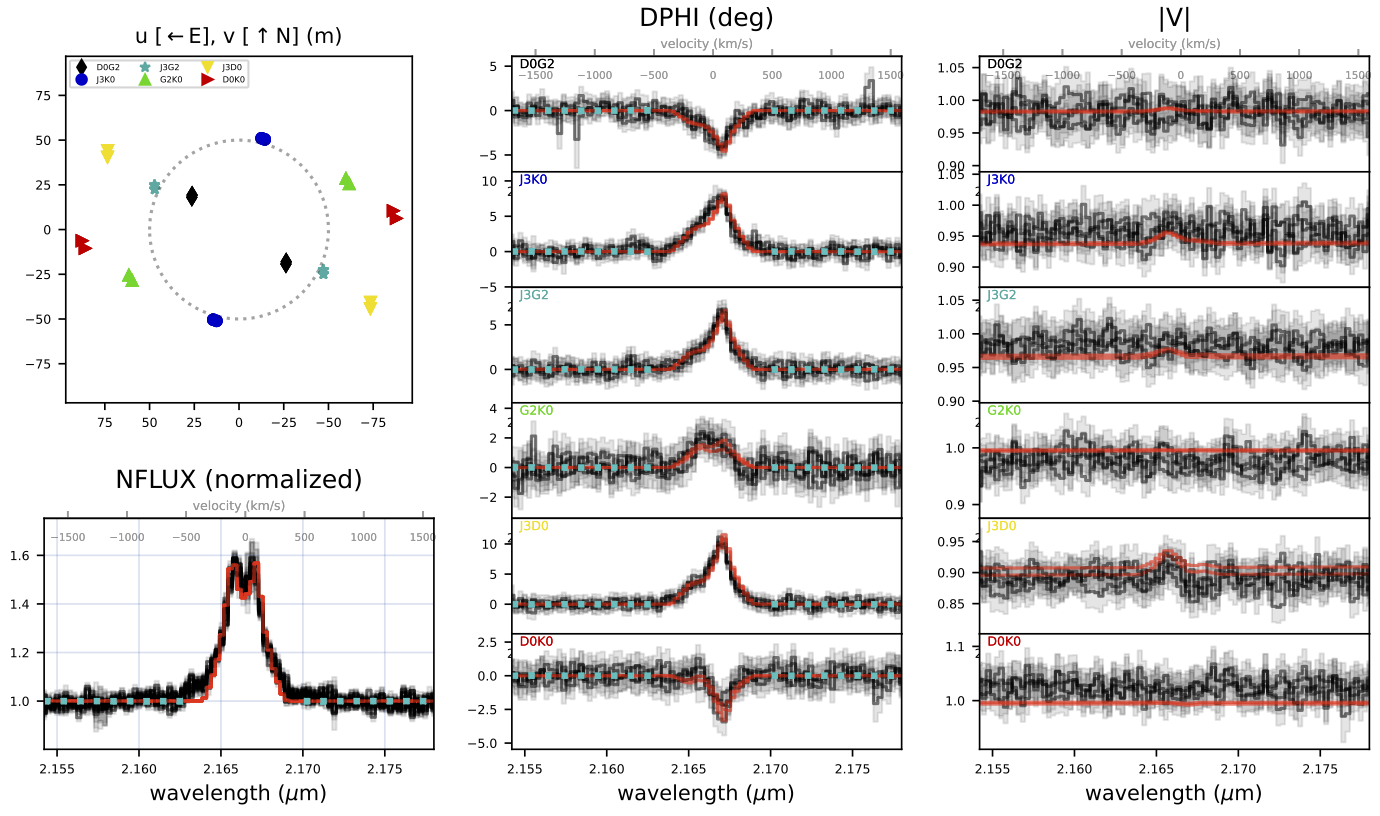}
      \caption{Fits to the second GRAVITY dataset from a model formed of Keplerian disk and two point sources. The inset labels are the VLTI baselines of the observations. NFLUX, T3PHI, DPHI and |V| refer to the normalised flux, closure phase, differential phase and visibilities respectively.
              }
         \label{endkepfit}
   \end{figure*}

\begin{figure*}
   \centering
   \includegraphics[width=180mm]{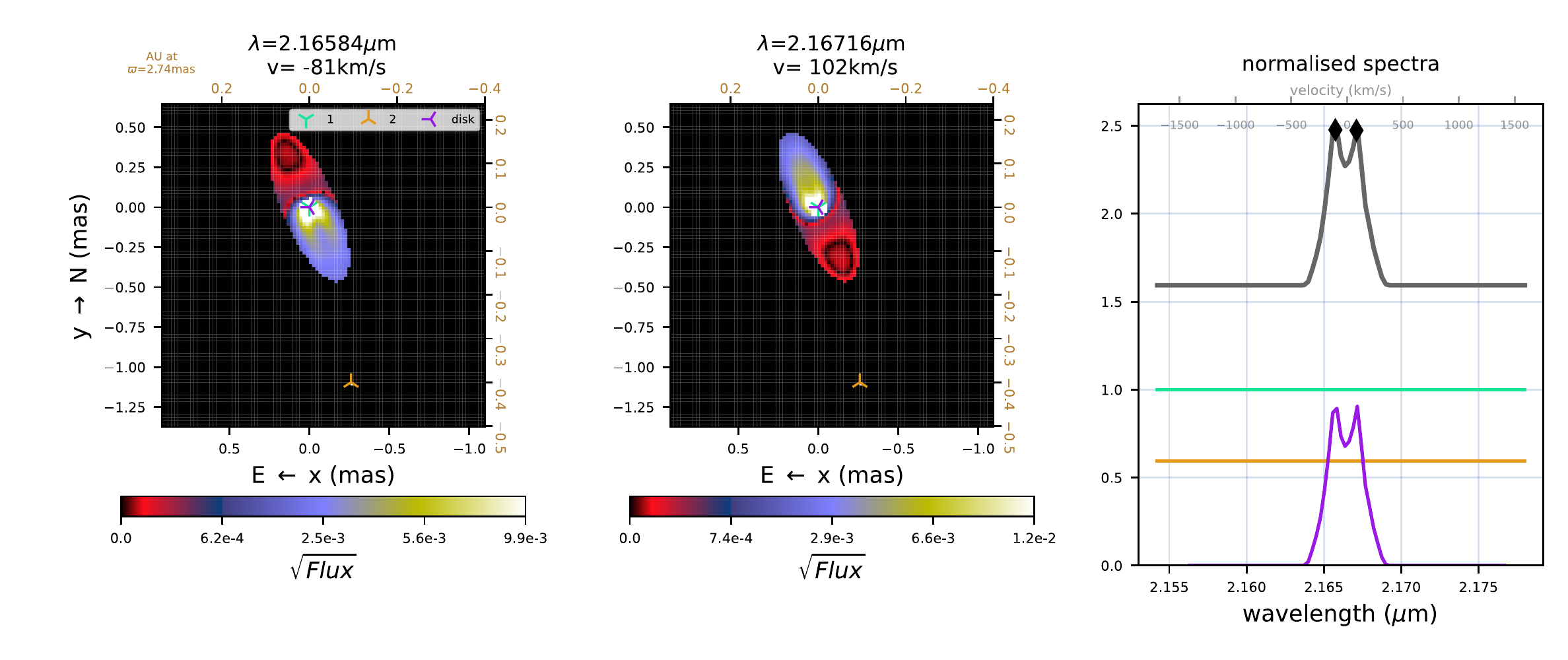}
      \caption{Model image and synthetic spectra for the toy model fit. The black line in the spectra is the total flux, the green line is the continuum flux of the brighter star, the orange line is the continuum flux of the dimmer star and the purple line is the disk contribution to the flux.
              }
         \label{endkepmod}
   \end{figure*}

We used this model to provide the starting parameters of another Keplerian disk fit. With these starting parameters, this Keplerian disk model could converge, resulting in the fits shown in Figures \ref{endkepfit} and \ref{endkepmod}. The Keplerian model improves the shape of the model differential phase and normalised flux profiles further and is the only model that additionally fits the small peak feature in the visibilities. Again, the main caveat is that the size of the disk as determined through the fitting process is too small to be detected with the VLTI. With only two epochs the measure of the velocity structure of the disk is also likely unreliable.

  \begin{table*}
\caption{Parameters derived from the GRAVITY fits of the final model. $\rho$ is the separation and PA is the position angle. The position of the primary was fixed at (0,0). In this best-fitting model both stars are modelled as uniform disks with 0.2\,mas diameters, such that they appear as point sources. $f_K$ is the flux ratio of the dimmer star in the $K$ band in reference to the flux of the normalised brightest star (fixed to 1 during the fitting process). $f_{\textnormal{line}}$ is the flux of the Br$\gamma$ line with respect to the continuum, $w_{\textnormal{line}}$ is the full-width at half maximum (FWHM) of the line} and $\lambda_{\textnormal{line}}$ is its wavelength.         
\label{params}      
\centering  
\begin{tabular}{c c c c c c c c}          
\hline\hline
Mean Julian Date & $\chi^{2}_{\textnormal{red}}$ & $f_\mathrm{K}$ & $\rho$ & PA & $f_{\textnormal{line}}$ & $w_{\textnormal{line}}$ & $\lambda_{\textnormal{line}}$ \\
  (MJD) &        &        & (mas) & ($^{\degr}$) & & (nm) & ($\mu$m) \\
\hline
59463.117 & 1.243 & 0.599$\pm$0.017 & 1.14$\pm$0.02 & 58.1$\pm$1 & 1.06$\pm$0.01 & 2.422$\pm$0.006 & 2.1661055$\pm$0.000002 \\
59476.031 & 1.321 & 0.516$\pm$0.019 & 1.16$\pm$0.03 & $-$167$\pm$0.8 & 0.892$\pm$0.001 & 2.643$\pm$0.004 & 2.166367$\pm$0.000001 \\
\hline
\end{tabular}
\end{table*}


Whilst the exact physical nature of the disk cannot be determined with our limited datasets, we conclude from our GRAVITY data 1) that the system is indeed a binary, as all the best-fitting models include two sources, 2) that the separation of these two sources is $\sim$1\,mas (across all models), and 3) that the brighter star in the system is likely to be the Be star, given the convergence on a model with the emission line in the brighter star and the successful (preliminary) detection of a disk around this brighter star. In the main body of the text we present the simplest binary model (with Br$\gamma$ emission in the brightest star) as our final model, given the uncertainty on the physical properties of this small disk that can be obtained with our dataset. Figures  \ref{endbinfit} and \ref{endbinmod} shown in the main text correspond to this model. Table \ref{params} shows the details of the final model.

   \section{Orbital details} \label{appendix2}
   
      \begin{figure*}
   \centering
   \includegraphics[angle=-90,width=145mm]{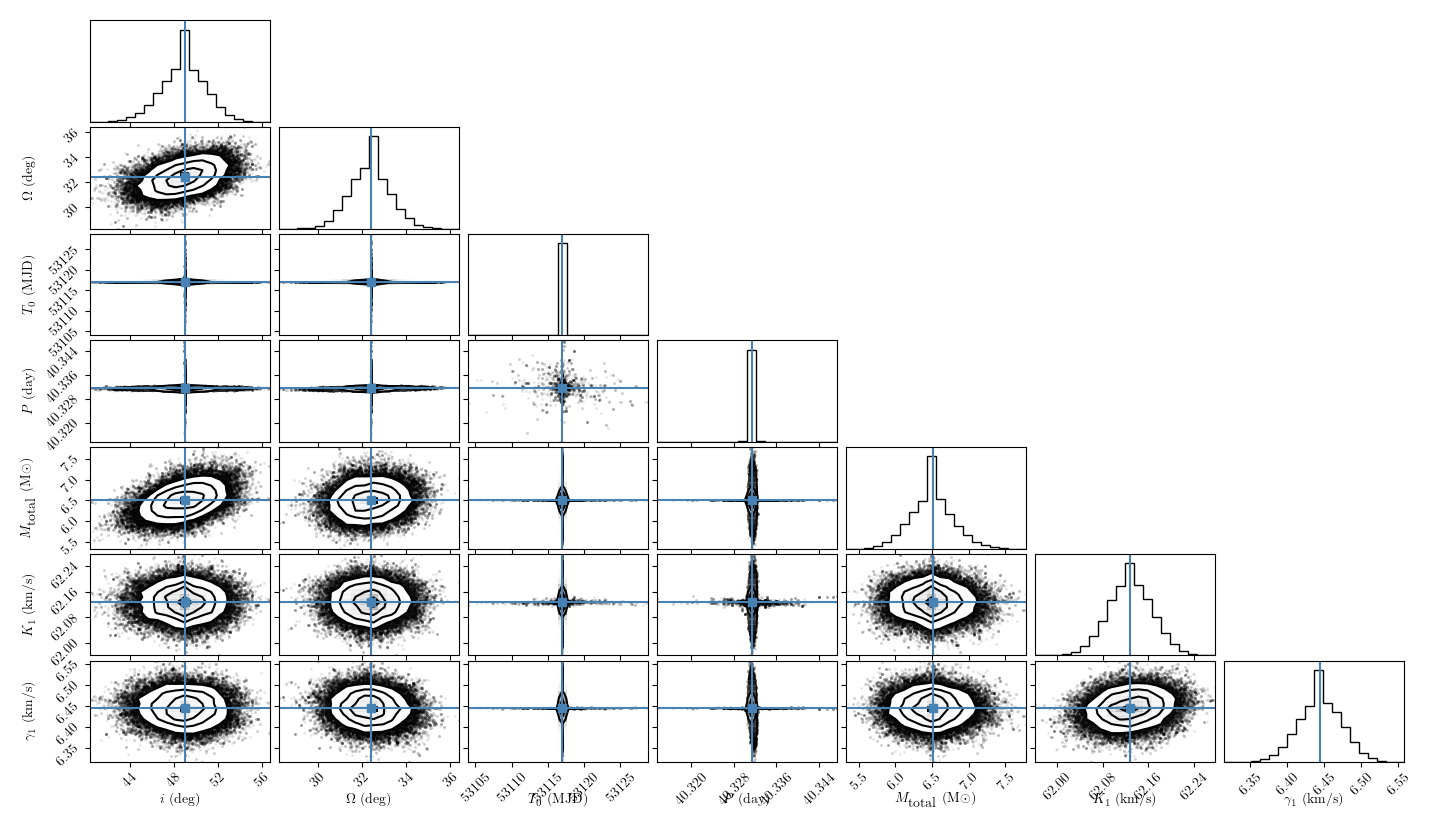}
      \caption{MCMC plot showing the error determination for the orbit shown in Fig.~ \ref{orbit}.
              }
         \label{mcmc}
   \end{figure*}
   
     \begin{table*}
\caption{Parameters of the orbit determined by \citet{juliahr6819} compared to the current derivations for different assumed distances $D$.  Circular orbits have been assumed in both works. MCMC was used to estimate the uncertainties.}           
\label{orbitparams}      
\centering  
\begin{tabular}{l l l l l}          
 \hline \hline
Parameter & \citet{juliahr6819} & \multicolumn{3}{c}{This work} \\ 
\hline 
$P_\mathrm{orb}$\,[d]          & $40.335 \pm 0.007$   & \multicolumn{3}{c}{$40.3315 \pm 0.0003$ } \\
$T_0$\,[MJD] ($\Phi=0$)        & $53116.9 \pm 1.1$    & \multicolumn{3}{c}{$53116.918 \pm 0.005$} \\
$\gamma$\,[\kms{}]             & $9.13 \pm 0.78$      & \multicolumn{3}{c}{$6.44 \pm 0.03$      } \\
$K_1$\,[\kms{}]                & $60.4 \pm 1.0$       & \multicolumn{3}{c}{$62.13 \pm 0.04$     } \\
$\Omega$ (\degr)               & {\it n/a}            & \multicolumn{3}{c}{$32.4 \pm 0.9$       } \\
$i$\,[deg]                     & $\sim32$ (computed)  & \multicolumn{3}{c}{$49.0 \pm 1.9$       } \\
$D$ (pc) [fixed]               & 340                  & 340            & 364            &  258  \\
$M_\mathrm{tot}$\,[$M_\odot$]  & $\sim7.5$ (computed) & $6.5 \pm 0.3$  & $8.0 \pm 0.4$  &  $2.8\pm 0.1$ \\ 
$K_2$\,[\kms{}]                & $4.0 \pm 0.8$        & $\sim25$ (computed)  &  $\sim32$ (computed)  &  $\sim4$ (computed)  \\
rms$_\mathrm{RV}$ (\kms)       &                      & \multicolumn{3}{c}{4.7}\\
rms$_\mathrm{as}$ ($\mu$as)    & {\it n/a}            & \multicolumn{3}{c}{11} \\
\hline
\end{tabular}
\end{table*}

%
%

\end{document}